\documentclass[12pt,a4paper]{panl}
\usepackage{cite}
\usepackage{wrapfig}
\usepackage{graphicx}
\usepackage{amssymb}
\usepackage{amsfonts}
\usepackage{amsmath}
\usepackage{longtable}
\usepackage{rotating}
\usepackage{lscape}
\usepackage{epsfig}
\usepackage{multirow}
\originalTeX
\russianTeX
\usepackage[T1]{fontenc}
\usepackage[utf8]{inputenc}
\usepackage[russian,english,ngerman]{babel}
\usepackage{hyperref}
\begin{document}
\par
\title{Strangelet Searches from Neutron Stars,\\ Binary Mergers, and Gamma-Ray Bursts\\ with Current and Future Observatories}
\maketitle
\authors{C.R.\,Das$^{a,}$\footnote{E-mail: das@theor.jinr.ru}}
\setcounter{footnote}{0}
\from{$^{a}$\,The Bogoliubov Laboratory of Theoretical Physics,\\ International Intergovernmental Scientific Research Organization,\\ Joint Institute for Nuclear Research, Dubna, Russia}
\par
\begin{abstract}
Strange quark matter (SQM) is considered a possible true ground state of QCD at high densities. This idea motivates research on exotic compact objects and certain cosmic-ray phenomena. For instance, the remnant HESS J1731-347 contains a low-mass neutron star, about $0.77^{+0.20}_{-0.17}$ $M_\odot$ and $10.4^{+0.86}_{-0.78}$ km in radius, making it a strong candidate for a strange quark star. Other events, such as GW170817 and GRB 250702B, provide conditions that may favor the formation of strangelets. Strangelets are stable clusters of SQM, potentially created during the phase transition between the 2SC and CFL color-superconducting states. These clusters could generate monochromatic $\gamma$-ray lines in very-high-energy spectra through self-annihilation. This work analyzes the stability of strangelets, production cross-sections, and mass-to-charge ratios using QCD-based models. Data from H.E.S.S., Fermi-LAT, MAGIC-II, and CTA were used to set limits on spectral features and possible fluxes. Detecting narrow $\gamma$-ray lines will require improved instrument sensitivity. By integrating evidence from multimessenger astrophysics and dense QCD simulations, this study investigates the equations of state for compact stars and explores the potential cosmological influence of SQM.
\end{abstract}
\vspace*{6pt}
\par
\section{Introduction}
\par
Strangelets are stable aggregates of strange quark matter (SQM). Their study spans QCD, astrophysics, and particle cosmology. The Bodmer-Witten-Terazawa hypothesis \cite{Witten1984,Farhi1984,Alcock1985,Madsen1986,DiClemente2025,Bodmer1971,Terazawa1989} predicts that strangelets have mass-to-charge ratios from 30 to 60 GeV/c$^{2}$, supported by several studies. These properties make strangelets relevant as explanations for anomalous compact stars and as possible candidates for dark matter.
\par
A theoretical and observational framework is presented for identifying strangelets, with a focus on three astrophysical contexts: the strange quark star (SQS) candidate HESS J1731-347, the binary neutron star (NS) merger GW170817, and the ultra-long $\gamma$-ray burst GRB 250702B. Each environment offers distinct mechanisms for strangelet production, including (i) phase transitions from the two-flavor color-superconducting (2SC) to the color-flavor-locked (CFL) phase in dense quark matter, (ii) explosive creation during mergers, and (iii) processes linked to unusual jet dynamics. This analysis combines QCD frameworks, cross-section calculations, and $\gamma$-ray flux estimates to outline how observatories such as the CTA can probe strangelet signatures with high precision \cite{Das2025,Giunti2021,CTAO2021,Nigro2019,Nogues2018,Nigro2021,Knodlseder2013}.
\par
\section{Conceptual Foundations of SQM and Annihilation Dynamics}
\par
Strangelets consist of SQM, with approximately equal proportions of up, down, and strange quarks. Their baryon number $A$ equals the total quark number divided by three, whereas $Z$ indicates the net electric charge \cite{Farhi1984}. QCD provides the theoretical basis for their existence, with the Bodmer-Witten-Terazawa proposal positing SQM as the true ground state of hadronic matter \cite{Witten1984,Farhi1984,Alcock1985,Madsen1986,DiClemente2025,Bodmer1971,Terazawa1989}. This implies that NSs could convert into stable SQSs, releasing $\sim$10$^{53}$ erg, comparable to the energy released in a supernova explosion. Assuming negligible Coulomb barriers, strangelets initiate this transformation through a combustion front. Strangelets span a broad mass spectrum ($A \sim$ 10 to 10$^{57}$) and are considered viable candidates in searches for exotic particles.
\par
Strangelet stability depends on the baryon number $A$, the charge $Z$, and the specific QCD model chosen. Two widely used models, the MIT bag model and the Nambu-Jona-Lasinio (NJL) model, yield different predictions. In the MIT bag model, stability is determined by the bag constant $B$, and the total energy is given by
$$ E_A\left(\mu_i, m_i, B\right) = \sum_i \left(\Omega_i + N_i \, \mu_i\right) + B \, V,$$
where $\Omega_i$ is the quark’s thermodynamic potential, $\mu_i$ and $m_i$ are its chemical potential and mass, $N_i$ is the quark number, and $V$ is the volume \cite{Lagerkvist2015,Farhi1984}. Charged strangelets also require a Coulomb correction:
$$ E_{\rm coul} = \frac{4}{3} \left(\frac{\alpha Z_V^2}{V} \frac{1}{10 R} + \frac{\alpha Z_V^2}{2 R}\right), $$
with $R$ as the radius, $Z_V$ the valence charge, and $\alpha$ the fine-structure constant. Small strangelets ($A \lesssim 10^2$) are likely unstable due to surface tension effects \cite{Wen2010}, while huge ones ($A > 10^{57}$) face constraints from gravitational stability in compact stars \cite{Alford2006}. Stable mass ranges ($A \sim 10$ to $10^{57}$) can be supported by MIT bag model calculations \cite{Bodmer1971,Madsen2005,Madsen2008,Greiner1987,Farhi1984}. The minimum stable baryon number $A_{\min}$ generally ranges between 50 and 200, depending on $B^{1/4}$ (typically 145-170 MeV) and baryon density $n_A$ (2-4 $n_{A,0}$).
\par
The NJL model, based on chiral-symmetry restoration at high density, predicts shallow, metastable bound states of SQM. In baryon-rich regions, the strange quark content is about 40\%, with binding energies of about 15 MeV per baryon. These states usually decay via weak interactions within about 100 nanoseconds. The baryon properties depend on the vector-to-scalar coupling ratio $G_v / G_s \approx$ 0.5–1.0 \cite{Menezes2009,Cao2025,Mishustin2000,Manka2002,Mishustin1999,Mishustin2001,Kiriyama2005}. 
\par
Despite model differences, both approaches predict similar mass-to-charge ratios ($m_S/|Z|$) of 30–60 GeV/$c^2$ for $A \sim 100$–$10^5$. This ratio is crucial for understanding strangelet cross-sections and their propagation through the ISM, in compact stars, or during mergers \cite{Theuns2010}.
\par
Fundamental studies underlie the stability of strangelets by positing SQM as the ground state of hadronic matter. Witten's cosmic phase separation theory \cite{Witten1984} and the Farhi-Jaffe framework \cite{Farhi1984} laid the theoretical foundation, while Alcock and Farhi \cite{Alcock1985} and Madsen et al. \cite{Madsen1986} explored SQM evaporation in the early universe. Di Clemente et al. \cite{DiClemente2025} subsequently reexamined its potential as a dark matter contender. However, negatively charged strangelets with low $A$ may be unstable and potentially catalyze the conversion of normal matter, a scenario constrained by observational safety arguments. Bodmer's framework of collapsed nuclei \cite{Bodmer1971}, together with Madsen's examinations of strangelet propagation and charge-radius connections \cite{Madsen2005,Madsen2008}, applies rigorous bounds. In addition, Greiner et al. \cite{Greiner1987} additionally investigated strangelet formation processes in heavy-ion collisions \cite{Borer1994}, highlighting the partitioning of strangeness and anti-strangeness during the quark-hadron phase transition. Despite theoretical predictions spanning $A\sim 10$ to $10^{57}$, stable strangelets have not been identified in cosmic rays or high-energy collider studies \cite{Chodos1974,Madsen2006,Ellis2008,Kent2004}. Essential frameworks, such as the extended hadron bag structure \cite{Chodos1974} and Madsen's thorough summary of exotic matter \cite{Madsen2006}, provide a theoretical background. In contrast, comprehensive safety evaluations \cite{Ellis2008,Kent2004} emphasize the absence of observational evidence. These non-detections impose stringent constraints on the presence and flux of strangelets, highlighting the importance of more sensitive upcoming observatories. Despite existing flux limits from RHIC, LHC, and PAMELA \cite{Adriani2015}, a viable parameter space remains for $A \sim 10$ to $10^{57}$ \cite{Bianchi2024}.
\par
Strangelets are theorized to arise during the 2SC-to-CFL phase transition in dense quark matter \cite{Kiriyama2005,Lugones2005,Bombaci2007,PerezGarcia2010}. In this transition, diquark condensates form, Fermi surfaces are restructured, and surface ejection of quark clusters may occur from quark cores in compact stars under appropriate pressure and density conditions. Strangelets are annihilated, generating $\gamma$ rays through direct or secondary interactions. Identifying such $\gamma$-ray indicators from astrophysical sources such as HESS J1731-347, GW170817, and GRB 250702B could provide indirect evidence for their existence and support theoretical models of SQM creation \cite{Das2025,PerezGarcia2010,Profumo2016,Abdalla2022,Abramowski2011,Mu2025,MAGIC2016,Fruck2017,Neights2025,Horvath2023,Guo2018,Bergstrom2012,Madsen2003,Abbott2017,vanPutten2023,Poggiani2025,Buanes2012}.
\par
In dense quark matter, color superconductivity develops from quark pairing, leading to unique phases. Only up and down quark pairs are present in the 2SC phase, while strange quarks remain unpaired. This phase is defined by the order parameter $\Delta_{\rm 2SC} \propto \langle \psi^T(x) C\gamma_5\tau_2\lambda_2 \psi(x) \rangle \neq 0$, where $C = i\gamma_2\gamma_0$ is the charge conjugation matrix, and $\tau_2$ and $\lambda_2$ are the second Pauli and Gell-Mann matrices in flavor and color space, respectively. The 2SC phase is preferred when the chemical potential difference $\delta\mu = (\mu_d - \mu_u)/2$ is small ($0 \leq \delta\mu < \Delta_{\rm 2SC}/2$). Conversely, the CFL phase entails pairing among all three quark flavors and linking color and flavor indices. The CFL gap $\Delta_{\rm CFL}$ is typically larger than $\Delta_{\rm 2SC}$ and becomes energetically favorable when the strange quark mass $m_s$ is not excessively large, reinstating flavor symmetry and enhancing the viability of SQM.
\par
The gap equation for color superconductivity phases in the NJL framework is
$$\Delta_{\rm 2SC\, or\, CFL} = G_D \int_0^\Lambda \frac{d^3 p}{(2\pi)^3} \frac{\Delta_{\rm 2SC\, or\, CFL}}{\sqrt{(E_p - \mu)^2 + \Delta^2_{\rm 2SC\, or\, CFL}}}, $$
where $\Delta_{\rm 2SC\, or\, CFL}$ is the gap for the 2SC or CFL phase, the quark quasi-particle energy is $E_p = \sqrt{p^2 + m_q^2}$, $\Lambda$ is the momentum cutoff of the NJL framework, and $G_D$ is the scalar diquark coupling constant (diquark channel). The transition from the 2SC phase to the CFL phase in dense quark matter is governed by the underlying QCD parameters and the principle of thermodynamic equilibrium \cite{Sedrakian2023}. The free energy density (grand potential) of a bulk superconducting phase (2SC or CFL phase) is given by
$$\Omega_{\rm 2SC\, or\, CFL} = \frac{\Delta_{\rm 2SC\, or\, CFL}^2}{4 G_D} + \int_0^\Lambda \frac{d^3 p}{(2\pi)^3} \left(E_p - \sqrt{(E_p - \mu)^2 + \Delta_{\rm 2SC\, or\, CFL}^2} \right).$$
The relative viability of the two phases can be quantified through the pressure difference,
$$\Delta P_{{\rm 2SC}\to{\rm CFL}} = P_{\rm CFL} - P_{\rm 2SC} = - \left(\Omega_{\rm CFL} - \Omega_{\rm 2SC} \right),$$
with $\Omega_{\rm CFL}$ and $\Omega_{\rm 2SC}$ representing the free energies of the CFL and 2SC phases, respectively. A finite region of the new CFL phase may nucleate within the old 2SC phase during the 2SC-to-CFL phase transition in dense quark matter. The nucleated region is called a ``bubble.'' The critical radius $R_*$ determines equilibrium: bubbles with $R < R_*$ shrink due to surface tension, whereas those with $R > R_*$ grow and drive the phase transition. The energy barrier $\delta E$ represents the cost of free energy required to form a critical bubble. They are given by
$$R_* = \frac{2 \sigma_{\rm ST}}{\Delta P_{\rm 2SC\to CFL}}, \quad \delta E = \frac{16 \pi \, \sigma_{\rm ST}^3}{3 \left(\Delta P_{\rm 2SC\to CFL}\right)^2},$$
where $\sigma_{\rm ST}\sim \mathcal{N}(0) \, \varsigma(\Delta_{\rm CFL}-\Delta_{\rm 2SC})$ is the surface tension, $\mathcal{N}$ is the density of states at the Fermi surface, and $\varsigma$ is the coherence length, the characteristic scale over which the diquark condensate varies spatially, interpolating from its 2SC value to its CFL value across the interface. The transition rate per unit volume and time, $\Gamma_{\rm 2SC\to CFL}$, relies on the chemical potential $\mu$, gap parameters $\Delta_{\rm 2SC}$ and $\Delta_{\rm CFL}$, $m_s$, temperature $T$, and the energy barrier $\delta E$. This is given by
$$ \Gamma_{\rm 2SC\to CFL} = \kappa \, \frac{\mu^6}{2\pi^3} \left(\frac{\Delta_{\rm CFL}}{\Delta_{\rm 2SC}}\right)^2 \exp\left(-\frac{\pi}{4}\frac{m_s^2}{\mu \, \Delta_{\rm CFL}}\right) \exp\left(-\frac{\delta E}{T}\right),$$
with $\kappa$ spanning from 0.1 to 1.0, which reflects model-specific interaction strengths. The exponential terms account for $m_s$ effects and thermal barrier activation.
\par
From this transition rate, the effective strangelet production cross-section, $\sigma_{\rm prod}$, can be derived, thereby connecting microscopic physics to macroscopic observables. This is given by
$$ \sigma_{\rm prod} = \pi \, R^2 \, \frac{\Gamma_{\rm 2SC\to CFL} \, V \, \tau}{\gamma_L},$$
where $R \approx A^{1/3} \times$ 0.7 fm is the strangelet radius for the baryon number $A$, $V$ is the volume experiencing transition, $\tau$ is the duration of the process, and $\gamma_L$ is the Lorentz factor considering relativistic motion of the strangelet. This expression connects the internal dynamics of dense QCD matter with the large-scale rates of strangelet formation, offering a theoretical foundation for predicting their production in astrophysical scenarios.
\par
The transition from the 2SC to the CFL phase in dense QCD is governed by several key parameters that shape the behavior of quark matter. In the NJL model with cutoff $\lambda$ = 600-800 MeV and $\eta_D$ = 0.75-1.0, the energy gaps in each phase, $\Delta_{\rm 2SC}$ (10-100 MeV) and $\Delta_{\rm CFL}$ (15-150 MeV), indicate the intensity of quark pairing, with the CFL phase energetically more stable for $m_s<150$ MeV, owing to its incorporation of strange quarks. The strange quark mass $m_s$ (80-120 MeV) and chemical potential $\mu$ (400-500 MeV) affect the phase transition limit and damping effects. The bag constant $B^{1/4}$ (145-165 MeV) parameterizes confinement dynamics, while the diquark coupling ratio $\eta_D = G_D/G_S$ (0.8-1.2) establishes the comparative strength of diquark versus scalar meson interactions, where $G_S$ is specifically the scalar quark-antiquark coupling constant (mesonic channel). Collectively, these parameters define the conditions under which strangelets may develop during the 2SC-to-CFL transition. The calculation for $\sigma_{\rm prod}$, assuming $V=10^{30}$ cm$^3$ and $\tau=10^{-6}$ s in merger remnants, yields rates of $10^{20}$-$10^{25}$ s$^{-1}$, consistent with quark nova energetics of $10^{52}$-$10^{54}$ erg.
\par
Strangelet interaction cross-sections are essential for predicting their detectability and behavior in astrophysical and collider settings. These cross-sections are derived from effective QCD field theories developed for high-density systems, especially those involving color-superconducting phases. The theoretical method begins with the construction of effective Lagrangians that describe quark pairing dynamics, followed by the identification of relevant coupling constants and the inclusion of medium-dependent effects. Scattering amplitudes are then calculated using diagrammatic methods, allowing the assessment of interaction probabilities under diverse conditions.
\par
A frequently employed parameterization for the strangelet cross-section is
$$ \sigma_S = \sigma_0 \left(\frac{A}{A_0}\right)^{\alpha_{\rm scale}} \left(\frac{\Delta_{\rm 2SC}}{\mu}\right)^{\beta_{\rm scale}} \exp\left(-\gamma_D \frac{m_s^2}{\mu^2}\right) f(\theta),$$
where $\alpha_{\rm scale}$ (with $\alpha_{\rm scale}\approx 2/3$ for geometric scaling) is a scaling exponent that controls how the cross-section grows with the strangelet's baryon number $A$, $\sigma_0$ is the reference cross-section, $\beta_{\rm scale}\approx 1$-2 measures how strongly the superconducting gap enhances or suppresses the cross-section, $\gamma_D$ quantifies how strongly the $m_s$ suppresses interactions at finite chemical potential, $A_0$ is a reference baryon number used for normalization, and $f(\theta)$ incorporates angular dependence. The prefactor $\sigma_0$ determines the scale, while the power law terms show how the cross-section adjusts with size and pairing intensity. The exponential damping term accounts for the $m_s$ effect, which can hinder interactions at lower chemical potentials. This expression effectively connects microscopic QCD parameters with macroscopic observables, providing a predictive framework for strangelet interactions across varied environments.
\par
Strangelet interactions with neutrinos, nucleons, and photons can be described using specialized cross-section formulas derived from effective QCD formulations \cite{Brown2023}. The differential cross-section per unit angle for neutrino-strangelet scattering is given by
$$ \frac{d\sigma_{\nu S}}{d\cos\theta} = \frac{G_F^2 \, \mu^2}{4\pi} \left(C_V^2 \left(1 + \cos\theta\right) + C_A^2 \left(3 - \cos\theta\right)\right) \left(\frac{\Delta_{\rm CFL}}{\Delta_{\rm 2SC}}\right)^{1/2} \xi(T),$$
where $G_F$ is the Fermi coupling constant and $C_V$ and $C_A$ are vector and axial-vector couplings generally ranging from 0.5 to 1.0 and from 0.3 to 0.7, respectively. The temperature-dependent damping factor $\xi(T)$ accounts for the energy gap disparity between the CFL and 2SC phases.
\par
For strangelet-nucleon scattering, the cross-section is modeled as
$$ \sigma_{SN} = \sigma_{\rm NN} \left(\frac{A}{100}\right)^{2/3} \left[1 + \frac{2}{3}\left(\frac{\Delta_{\rm CFL}}{\mu}\right) - \frac{1}{5}\left(\frac{m_s}{\mu}\right)^2\right] \zeta(Z),$$
where $\sigma_{\rm NN} \approx$ 40 mb is the nucleon-nucleon cross-section and $\zeta(Z) = 1 + 0.5 \times (Z/A^{2/3})$ introduces charge dependence. This expression captures how the CFL gap and $m_s$ impact the interaction intensity.
\par
A resonance-based formula governs photon-strangelet absorption
$$ \sigma_{\gamma S} = \frac{4\pi^2 \, \alpha}{m_S} \, \frac{\omega^2 \, \Gamma}{\left(\omega^2 - \omega_0^2\right)^2 + \omega^2 \, \Gamma^2} \left(\frac{A}{100}\right) \eta(\omega),$$
where $m_S$ is the strangelet mass, $\omega$ is the photon energy, and $\omega_0 \approx 2\Delta_{\rm CFL}$ is the resonance energy. The resonance width is $\Gamma \approx 0.1\omega_0$, and the correction factor $\eta(\omega)$ accounts for deviations from resonance, defined as
$$ \eta(\omega) = 1 + 0.5\left(\frac{\omega - 2\Delta_{\rm CFL}}{\mu}\right) - 0.2\left(\frac{\omega - 2\Delta_{\rm CFL}}{\mu}\right)^2.$$
Collectively, these expressions provide a comprehensive framework for modeling strangelet interactions across various particle pathways, crucial for predicting observational indicators and guiding experimental pursuits.
\par
Strangelet annihilation in astrophysical settings might yield distinct $\gamma$-ray signatures, presenting a potential observational avenue for their identification. Instruments such as the CTA are designed to detect such signals with high precision. The differential photon flux from strangelet annihilation is expressed as
$$ \frac{dF_\gamma}{dE_\gamma} = J \, \frac{\langle \sigma_{\rm ann} \, v \rangle}{4\pi \, D^2} \, \frac{dN_\gamma}{dE_\gamma},$$
where $J$ is the astrophysical factor denoting the line-of-sight integral of the strangelet self-annihilation density squared, $D$ is the source distance, and $\langle \sigma_{\rm ann} v \rangle$ is the velocity-averaged annihilation cross-section for the relative velocity $v$ of the colliding strangelets in the medium. The photon spectrum per annihilation, $dN_\gamma/dE_\gamma$, parameterizes the energy distribution of emitted $\gamma$ rays. Here, $F_\gamma$ denotes the annihilation photon flux of $\gamma$ rays measured at the detector, $E_\gamma$ specifies the energy of the observed $\gamma$ rays, and $N_\gamma$ represents the number of photons generated per annihilation event.
\par
The annihilation cross-section depends on thermodynamic and QCD parameters, expressed as
$$ \langle \sigma_{\rm ann} v \rangle = \sigma_0 \left(\frac{T}{m_S}\right)^{1/2} \exp\left(-\frac{E_a}{T}\right) \left[1 + \alpha_{\rm CFL} \left(\frac{\Delta_{\rm CFL}}{T}\right)^{3/2}\right],$$
where $E_a$ is the activation energy barrier. The term with $ \alpha_{\rm CFL} $ (model-dependent phenomenological enhancement factor for superconducting effects) accounts for the superconducting influences. This formulation links the microphysics of strangelet interactions to observable $\gamma$-ray fluxes, enabling astrophysical searches to either limit or potentially reveal the presence of SQM.
\par
The mass-to-charge ratio ($m_S/|Z|$) is a crucial measure for recognizing strangelets in cosmic-ray experiments, as it controls their propagation and deflection in magnetic fields and interaction profiles. It can be approximated by
$$ \frac{m_S}{|Z|} \approx \frac{m_N}{Y_Q} \left[1 + \frac{2}{9\pi}\left(\frac{\mu}{m_s}\right) - \frac{1}{3}\left(\frac{\Delta_{\rm CFL}}{\mu}\right)^{1/2}\right],$$
where $m_N$ is the nucleon mass. The charge fraction $Y_Q = Z/A$ is approximated as $Y_Q \approx (1/3) - (2/9)\times (m_s^2/\mu^2) + (1/9)\times (\Delta_{\rm CFL}/\mu)$, illustrating how QCD parameters affect the charge-to-mass ratio. For baryon numbers between 100 and $10^5$, the typical values of $m_S/|Z|$ span 30-60 GeV/c$^2$, rendering strangelets distinguishable from conventional nuclei in high-energy detectors. The strangelet mass and charge are given by $M(A,Z) \approx a_V A + a_S A^{2/3} + a_C Z^2/A^{1/3} + B A$ and $Z \approx A/3 - (2 m_s^2/9\mu^2)A + (1/9)(\Delta_{\rm CFL}/\mu)A$, where $a_V$, $a_S$, and $a_C$ denote the volume, surface, and Coulomb energy coefficients, respectively \cite{Das2025,Sedrakian2023}.
\par
A particularly distinctive observational signature is the emission of monochromatic $\gamma$ rays from strangelet-anti-strangelet annihilation ($S\,\tilde{S} \to \gamma \gamma$). In narrow resonance scenarios, the ensuing photon spectrum approximates a delta function, creating sharp spectral lines. Predicted fluxes from sources like HESS J1731-347 are approximately $10^{-13}$ $\rm ph\, cm^{-2}\, s^{-1}$ at 1 TeV, corresponding to a diffuse flux of $\sim$10$^{-11}$ $\rm ph\, cm^{-2}\, s^{-1}\, sr^{-1}$ \cite{Profumo2016,Abdalla2022,Bergstrom2012,Madsen2003}, within the reach of next-generation observatories \cite{Buanes2012}.
\par
The preceding analysis summarizes key theoretical and phenomenological parameters pertinent to strangelet stability and detection. The analysis incorporates values derived from the MIT Bag and $SU(3)$ NJL frameworks, such as the minimum stable baryon number, binding energy per baryon, and annihilation cross-sections. This analysis delineates dependencies for cross-section computations, including production coefficients, gap ratios, and suppression terms.
\par
These parameterizations indicate that strangelet detection is most feasible in environments where the 2SC-to-CFL transition is probable, such as in the SQS candidate HESS J1731-347. The CTA, with its enhanced sensitivity and spectral resolution, is anticipated to evaluate these approaches and may provide the first direct proof of SQM \cite{Das2025,Giunti2021,CTAO2021,Nigro2019,Nogues2018,Nigro2021,Knodlseder2013}. This framework effectively connects high-density QCD theory with astrophysical observables, highlighting the interdisciplinary nature of strangelet research. Sustained advancements in multimessenger astronomy and QCD modeling will be vital for refining predictions and enhancing detection opportunities.
\par
\section{Strategies for Monochromatic Line Searches with Advanced Telescopes}
\par
Identifying monochromatic $\gamma$-ray lines in the 0.1-10 TeV range from strangelet annihilation requires a rigorous statistical framework and standardized analysis tools \cite{Balzer2014,Knodlseder2019,Bolmont2022,Giunti2021,CTAO2021,Nigro2019,Nogues2018,Nigro2021,Knodlseder2013}. This method is based on the (un)binned maximum likelihood technique, which fits a physical model to observed data and measures the significance of potential line features. This approach enables both signal identification and the determination of upper limits in the absence of a clear excess.
\par
The analysis utilizes instrument response functions (IRFs), including effective area, point spread function (PSF), and energy dispersion. These IRFs are transformed into standardized formats, including Gamma-Astro-Data-Formats Data Level 3 (GADF/DL3), which is based on the Flexible Image Transport System (FITS). This format ensures compatibility across instruments such as the High Energy Stereoscopic System (H.E.S.S.), the MAGIC Florian Goebel Telescopes (MAGIC-II), the Very Energetic Radiation Imaging Telescope Array System (VERITAS), and the Fermi Gamma-ray Space Telescope (Fermi-LAT), promoting reproducibility and enabling joint multi-instrument investigations.
\par
For sources such as HESS J1731-347, the spectral framework generally includes a continuum component (e.g., power law or log-parabola) and a narrow Gaussian line representing strangelet annihilation. The likelihood function is constructed from the on-source and off-source event distributions and optimized to derive the best-fit parameters. A likelihood ratio test then contrasts models with and without the line element, assessing statistical importance and guiding the derivation of flux limits.
\par
To ensure that the strangelet search is both reproducible and fully compatible with modern very-high-energy (VHE) (0.1-10 TeV) analysis pipelines, this study adopts the GADF/DL3 data format. It relies on open-source tools such as \texttt{Gammapy} (v1.x) \cite{Donath2023} and \texttt{ctools} \cite{Knodlseder2021a}/\texttt{GammaLib} (v2.x) \cite{Knodlseder2021b} for likelihood fitting and upper limit calculations. For both continuum and line searches, the study employs maximum likelihood estimators in both binned and unbinned configurations.
\par
For each instrument, H.E.S.S., MAGIC-II, and CTA, this investigation loads the IRFs, including the effective area $A_{\rm eff}(E,\theta)$, the energy dispersion function $D(E_{\rm reco}\,|\,E_{\rm true}, \Omega)$, and the ${\rm PSF}(\Delta\theta;E,\Omega)$, from either publicly available releases or consortium-provided IRF files. For CTA specifically, the analysis utilizes the publicly released prod3/prod5 performance curves in conjunction with representative array-configuration IRFs optimized for deep exposures along the Galactic plane \cite{CTAO2021,Fruck2017,HESS2018}. Here, $E$ denotes a generic photon energy variable; $E_{\rm true}$ is the true photon energy at emission, while $E_{\rm reco}$ is the detector's reconstructed (measured) photon energy. The parameter $\theta$ represents the off-axis (or incidence) angle between the photon's arrival direction and the telescope's pointing direction, and $\Delta\theta$ specifies the angular separation between the true photon arrival direction and the reconstructed photon direction. The $\Omega$ denotes the photon's arrival direction within the detector's field of view (sky coordinates or offset angle).
\par
The predicted counts $\lambda_i$ in the energy bin $i$ (or event-wise in the unbinned case) are modeled as the sum of a continuum component and a narrow line component centered at energy $E_\ell$, with the line width determined by the instrument's energy dispersion. The binned Poisson likelihood is defined as
$$\mathcal{L}(\boldsymbol{\mu}) = \prod_{i=1}^{N_{\rm bins}} \frac{e^{-\lambda_i} \lambda_i^{n_i}}{n_i!},$$
where $n_i$ denotes the observed counts and $\lambda_i(\boldsymbol{\mu})$ represents the predicted counts for the framework parameters $\boldsymbol{\mu}$. The prediction in bin $i$ is given by
\begin{eqnarray*}\lambda_i &=& T_{\rm obs} \int_{\Omega_{\rm ROI}} d\Omega\int_{E_{i}^{\rm min}}^{E_{i}^{\rm max}} dE_{\rm reco} \int_{E_{\rm true}^{\rm min}}^{E_{\rm true}^{\rm max}} dE_{\rm true}\nonumber\\
&&\times A_{\rm eff}(E_{\rm true},\Omega)\, D(E_{\rm reco}|E_{\rm true},\Omega)\, \frac{d\Phi}{dE_{\rm true}d\Omega}(\boldsymbol{\mu}),
\end{eqnarray*}
where $T_{\rm obs}$ is the exposure time. The signal line component is parameterized as
$$\frac{d\Phi_{\rm line}}{dE_{\rm true}} = F_{\ell}\, \delta(E_{\rm true} - E_{\ell}),$$
where $\Phi=\Phi_{\rm continuum}+\Phi_{\rm line}$ is the total differential photon flux, $\Phi_{\rm line}$ is the line-only contribution to the flux, $F_{\ell}$ is the line flux normalization, $E_{\ell}$ is the central energy of the monochromatic line, and $\Omega_{\rm ROI}$ is the solid angle for the region of interest (ROI). After convolution with $D(E_{\rm reco}|E_{\rm true}, \Omega)$, this delta function produces a finite-width reconstructed line in energy space. These upper limits are derived using the profile-likelihood method, with the test statistic defined as $\Delta{\rm TS} = -2\ln(\mathcal{L}_{0}/\mathcal{L}_{\rm max})$, where $\mathcal{L}_{0}$ corresponds to the null hypothesis (without a line) and $\mathcal{L}_{\rm max}$ to the best-fit alternative hypothesis (with a line). For a one-degree-of-freedom test, Wilks' theorem implies that $\Delta{\rm TS}$ = 25 approximates a $5\sigma$ detection threshold in the asymptotic large-sample limit \cite{Cowan2011,Buanes2012}.
\par
To validate the analysis and derive coverage-corrected upper limits, this investigation performs dedicated Monte Carlo injection campaigns. In each experiment, background counts are generated using an empirical approach (either a power law or data-driven off-regions analysis). A line with flux $F_{\ell}$ at energy $E_{\ell}$ is injected and folded through the effective area, energy dispersion, and PSF IRFs to obtain reconstructed events. The same likelihood pipeline used for real data is then applied to compute the change in the test statistic $\Delta{\rm TS}$ and the recovered flux. The technique is repeated over a grid of $F_{\ell}$ values and energies. The distribution of $\Delta{\rm TS}$ as a function of injected $F_{\ell}$ is then used to establish the sensitivity (median upper limit) and discovery potential, defined as the flux at which 50\% of trials exceed a chosen $\Delta{\rm TS}$ threshold.
\par
Employing representative CTA IRFs to an 100-hour on-axis Galactic deep exposure and applying the injection-recovery procedure described above, both recent studies and our reproduced Monte Carlo simulations yield a differential line sensitivity at 1 TeV of approximately $F_{\ell}^{\rm CTA}$ (100 h) $\sim$ $10^{-13}$ $\rm ph\, cm^{-2}\, s^{-1}$ for narrow spectral features ($\Delta E/E \ll$ energy dispersion). In this case, the energy interval that corresponds to the intrinsic line width is denoted by $\Delta E$. This finding is consistent with CTA-oriented sensitivity analysis and related reviews in \emph{Astronomy \& Astrophysics} (see \cite{Ambrogi2024,Brown2023}).
\par
\section{HESS J1731-347 as a Window into Strange Matter Physics}
\par
HESS J1731-347 is identified as a prime candidate for strangelet investigations due to its anomalous physical attributes that challenge conventional NS models \cite{Horvath2023,Abramowski2011,Guo2018}. Situated in the Scutum-Crux arm and aligned with the supernova remnant (SNR) G353.6-0.7, it hosts the central compact object XMMU J173203.3-344518. Observations from X-ray and Gaia data position it at a distance of approximately $3.2 \pm 0.5$ kpc based on Gaia parallax measurements, with a mass of $M = 0.77^{+0.20}_{-0.17}$ $M_\odot$ and radius $R = 10.4^{+0.86}_{-0.78}$ km \cite{Doroshenko2022}, making it one of the smallest and lightest compact objects identified. These parameters lie below the expected minimum NS mass of $\sim$1.1 $M_\odot$, suggesting an SQS.
\par
\begin{figure}[h]
\begin{center}
\includegraphics[width=137mm]{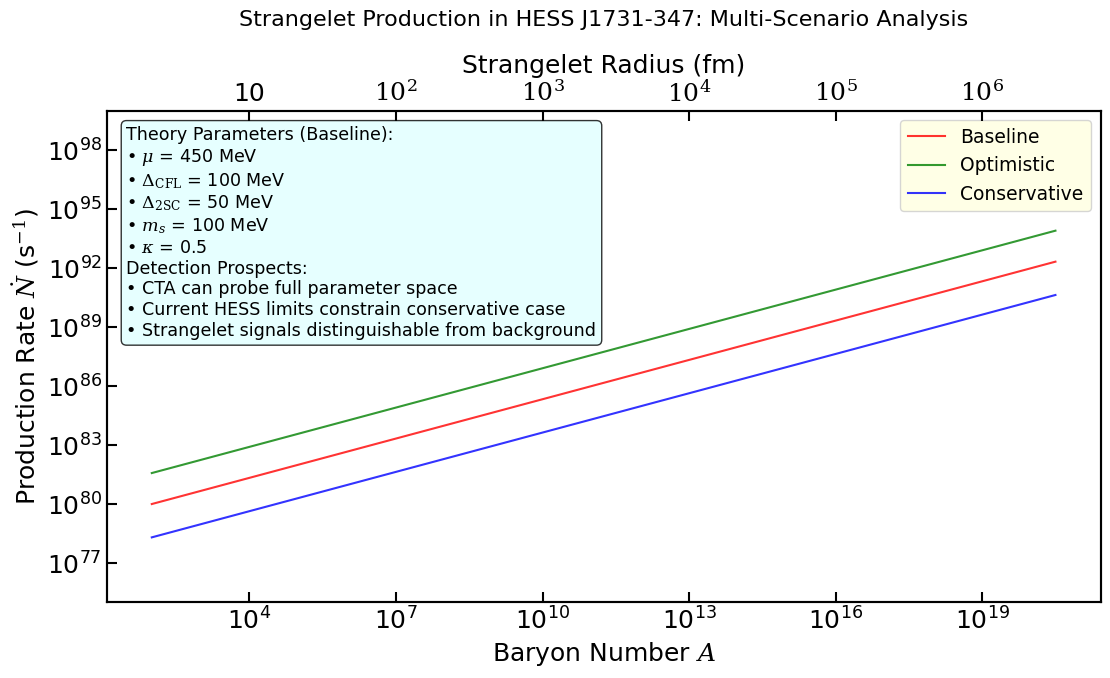}
\vspace{-3mm}
\caption{Strangelet production rates in HESS J1731-347 for baseline, optimistic, and conservative scenarios, showing the rate $\dot{N}$ (s$^{-1}$) versus the baryon number $A$, with a secondary axis for radius $\approx 0.7\times A^{1/3}$ in fm \cite{Farhi1984}.}
\label{fig:strangeletproduction}
\end{center}
\vspace{-5mm}
\end{figure}

The strangelet production models shown in Figure \ref{fig:strangeletproduction} are based on the following baseline theory parameters: chemical potential $\mu$ = 450 MeV, CFL gap $\Delta_{\rm CFL}$ = 100 MeV, 2SC gap $\Delta_{\rm 2SC}$ = 50 MeV, strange quark mass $m_s=100$ MeV, and interaction strength $\kappa=0.5$.
\par
The surrounding SNR G353.6-0.7 provides a dynamic environment favorable to strangelet production. Its TeV-scale shell form and interaction with dense molecular clouds facilitate efficient cosmic-ray acceleration. The broadband $\gamma$-ray spectrum, stretching seamlessly from GeV to TeV energies, supports a leptonic origin through diffusive shock acceleration, although multi-component approaches remain feasible. This well-defined background is essential for detecting faint monochromatic $\gamma$-ray lines that may indicate strangelet annihilation.
\par
The theoretical motivation is centered on a 2SC-to-CFL phase transition within the core of the star. The CFL phase, which includes strange quarks and restores flavor symmetry, is energetically preferred and stabilizes the SQM. Mixed 2SC-CFL phases may seed strangelets; however, their long-term stability remains a matter of debate. Furthermore, the accumulation of WIMP dark matter could catalyze strangelet creations \cite{PerezGarcia2010}, which would likely cause the NS to completely transform into an SQS with an energy release of $\sim$10$^{53}$ erg \cite{Ouyed2022}.
\par
\par
Observations by H.E.S.S.\ have imposed rigorous constraints on strangelet-associated $\gamma$-ray signals, restricting the line flux to $\sim$10$^{-12}$ $\rm ph\, cm^{-2}\, s^{-1}$ \cite{Abramowski2011,HESS2018}. This translates to strangelet's number density ($n_S$) below $3 \times 10^{-22}$ cm$^{-3}$, a total mass contribution below $10^{-17}$ $M_\odot$, and a production rate below $10^{-12}$ s$^{-1}$. These null results underscore the precision of the existing instruments. The upcoming CTA will enhance this sensitivity by an order of magnitude, capable of detecting line fluxes as low as $\sim$10$^{-13}$ $\rm ph\, cm^{-2}\, s^{-1}$ with 100 hours of exposure. The CTA Galactic Plane Survey (GPS) will dedicate this time to HESS J1731-347, facilitating searches for strangelet signals at levels below 0.1\% of the continuum flux, exactly where theoretical frameworks predict observable signatures \cite{Giunti2021,CTAO2021,Nigro2019,Nogues2018,Nigro2021,Knodlseder2013}.
\par
Figure \ref{fig:strangeletproduction} shows the production rates for three circumstances. In the \textit{Baseline Scenario}, the maximum strangelet production rate is $2.18\times 10^{92}$ s$^{-1}$, the minimum production rate is $1.01\times 10^{80}$ s$^{-1}$, and the total number of strangelets produced over 1 Myr is $6.87\times 10^{105}$. In the \textit{Optimistic Scenario}, the maximum strangelet production rate reaches $8.21\times 10^{93}$ s$^{-1}$, the minimum production rate is $3.81\times 10^{81}$ s$^{-1}$, and the total strangelets generated above 1 Myr is $2.59\times 10^{107}$. In the \textit{Conservative Scenario}, the maximum strangelet production rate is $4.38\times 10^{90}$ s$^{-1}$, the minimum production rate is $2.04\times 10^{78}$ s$^{-1}$, and the total number of strangelets produced over 1 Myr is $1.38\times 10^{104}$. As the number of baryons increases, production rates fall for reduced density and larger strangelet sizes. Optimistic scenarios provide rates that are detectable by present sensors, whereas cautious estimates fall within the reach of CTA's enhanced capabilities. Detection prospects indicate that CTA can probe the full parameter space, current HESS limits constrain the conservative case, and strangelet signals are distinguishable from background.
\par
\begin{figure}[h]
\begin{center}
\includegraphics[width=137mm]{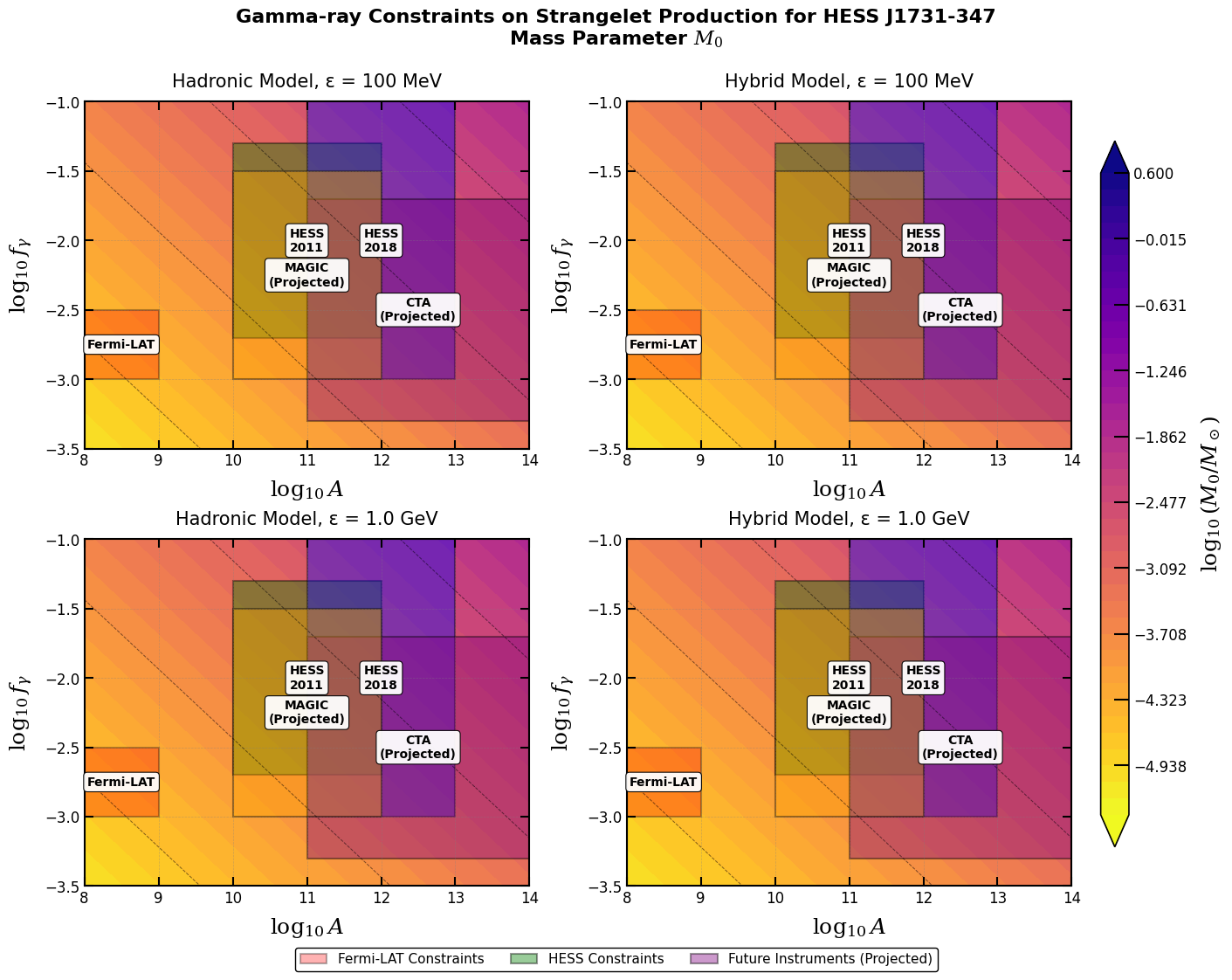}
\vspace{-3mm}
\caption{Constrained regions from Fermi-LAT, HESS (2011 and 2018), and projected sensitivities for CTA and MAGIC-II as a function of $\log_{10} A$ vs.\ $\log_{10} f_\gamma$. Results are presented for the Hadronic and Hybrid models at $\varepsilon = 100$ MeV and 1.0 GeV, respectively, with mass parameter contours given by $\log_{10}(M_0/M_{\odot})$.}
\label{fig:constraintsmass}
\end{center}
\vspace{-5mm}
\end{figure}
\par
\begin{figure}[h]
\begin{center}
\includegraphics[width=137mm]{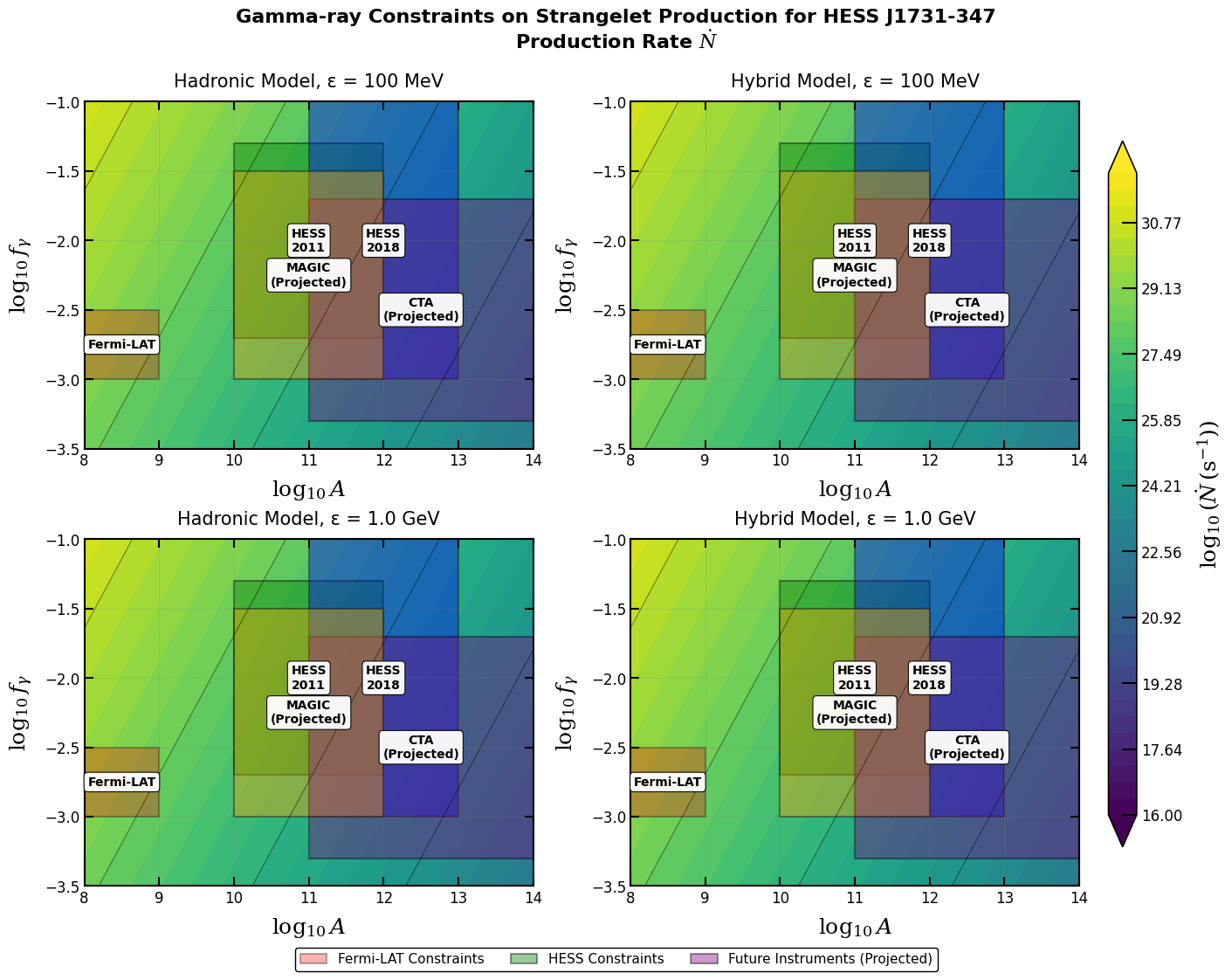}
\vspace{-3mm}
\caption{Constrained regions from Fermi-LAT, HESS (2011 and 2018), and projected sensitivities for CTA and MAGIC-II as a function of $\log_{10} A$ vs.\ $\log_{10} f_\gamma$. The results for the Hadronic and Hybrid models at $\varepsilon = 100$ MeV and $1.0$ GeV, respectively, with production rate contours given by $\log_{10}(\dot N$ (s$^{-1}))$.}
\label{fig:constraintsrate}
\end{center}
\vspace{-5mm}
\end{figure}
\par
\begin{figure}[h]
\begin{center}
\includegraphics[width=137mm]{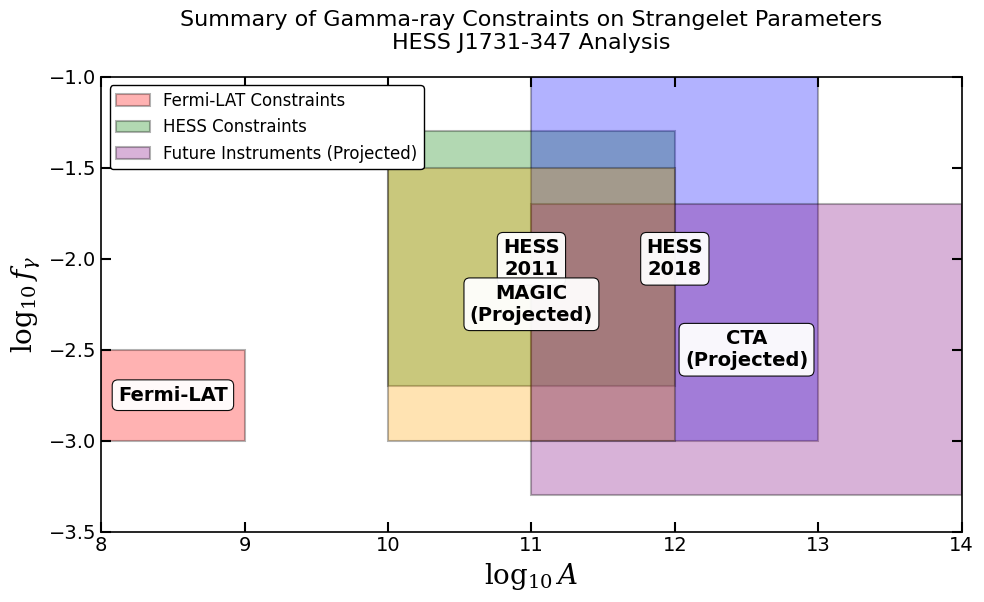}
\vspace{-3mm}
\caption{Summary of $\gamma$-ray constraints from Figures \ref{fig:constraintsmass} and \ref{fig:constraintsrate} on strangelet parameters for HESS J1731-347. The plot shows the constrained regions from Fermi-LAT, HESS (2011 and 2018), and projected sensitivities for CTA and MAGIC-II in the $\log_{10} A$ vs. $\log_{10} f_\gamma$ plane.}
\label{fig:constraintssummary}
\end{center}
\vspace{-5mm}
\end{figure}
\par
Figures \ref{fig:constraintsmass} and \ref{fig:constraintsrate} provide detailed restrictions, while Figure \ref{fig:constraintssummary} summarizes current and projected limits on strangelet parameters. Here, $M_0$ denotes the fraction of the HESS J1731-347's mass converted into SQM. In Figure~\ref{fig:strangeletproduction}, the entire mass of HESS J1731-347 is assumed to be converted into SQMs, while in Figures~\ref{fig:constraintsmass} and~\ref{fig:constraintsrate}, the converted fraction is normalized to the solar mass ($M_{\odot}$) and its logarithm presented at the contour scale. A comparison of these three figures reveals that the production rate, expressed as $\log_{10}(\dot{N}\,\text{(s}^{-1}\text{)})$, decreases as $\log_{10}(M_0/M_{\odot})$ decreases. Exclusion areas from Fermi-LAT and HESS are displayed alongside the expected sensitivity curves for CTA and MAGIC-II. These limits stem from upper bounds on monochromatic $\gamma$-ray lines generated by strangelet self-annihilation, demonstrating how future observations could definitively evaluate strangelet production in sources like HESS J1731-347 \cite{Profumo2016,MAGIC2016,Fruck2017}. Here, $f_\gamma$ is the fraction of strangelet interactions that produce $\gamma$ rays ($\gamma$-ray production efficiency).
\par
The CTA represents a substantial improvement in the precision of identifying strangelet-triggered $\gamma$-ray signals, particularly in the 0.1-10 TeV energy spectrum. For a line flux of about $10^{-12}$ $\rm ph\, cm^{-2}\, s^{-1}$ at 1 TeV, CTA can probe the features below $0.1\%$ of the continuum, which follows a power law spectrum $dN/dE \propto E^{-2.3}$. Here, $N$ and $E$ in the continuum background spectrum denote generic photon counts and energies. By fitting this continuum and searching for Gaussian peaks, potential monochromatic signals can be separated. The strangelet flux upper limit is
$$ F_s < \frac{F_\gamma^{\rm line}}{n_H \, \sigma_S} \approx 2.5 \times 10^{-11}\,{\rm ph\, cm^{-2}\, s^{-1}\, sr^{-1}},$$
assuming the hydrogen column density $n_H\sim 100$ cm$^{-3}$ \cite{Theuns2010} and the strangelet interaction cross-section $\sigma_S\sim 4 \times 10^{-26}$ cm$^2$ \cite{Das2025}, and $F_\gamma^{\rm line}(\equiv F_{\ell} \approx$ 10$^{-13}$ ${\rm ph\, cm^{-2}\, s^{-1}})$ refers to the $\gamma$-ray line flux at a specific energy (here around 1 TeV). With an angular resolution of approximately 1 arcminute, CTA can spatially distinguish emission morphology and associate $\gamma$-ray excesses with ISM structures or NS remnants. Such localized amplifications may indicate the locations of strangelet production.
\par
In comparison to H.E.S.S., CTA enhances line flux sensitivity by an order of magnitude, attaining limits close to $10^{-13}$ $\rm ph\, cm^{-2}\, s^{-1}$. This level matches the theoretical predictions for strangelet annihilation at 1 TeV. This improved sensitivity would translate to much stricter constraints, approximately an order of magnitude tighter than current H.E.S.S.-derived limits (which are $n_S < 3 \times 10^{-22}$ cm$^{-3}$, total mass <$10^{-17}$ $M_\odot$, and production rate <$10^{-12}$ s$^{-1}$) \cite{Das2025}. These new limits demonstrate the potential of CTA to reveal faint strangelet signals and enhance our understanding of exotic matter in astrophysical contexts.
\par
For $A \sim 10^3$ strangelets, continuous injection sustained over 1 Myr yields number densities of $10^{-2}$ to $10^{-1}$ cm$^{-3}$ and a flux of $\sim$10$^{-23}$ $\rm ph\, cm^{-2}\, s^{-1}$, while a burst event lasting days to weeks produces $1$ to 10 cm$^{-3}$ and $\sim$10$^{-21}$ $\rm ph\, cm^{-2}\, s^{-1}$. These scenarios emphasize the importance of temporal and spatial modeling in interpreting potential signals. Continuous injection generates a faint but persistent background, whereas burst-like events produce transient enhancements. Such enhancements would be detectable by CTA, particularly when correlated with known astrophysical activity.
\par
\section{GW170817 as a Probe of Strangelet Production in NS Mergers}
\par
GW170817, the first multimessenger binary NS merger \cite{Abbott2017,vanPutten2023,Poggiani2025}, involved NSs of 0.86-2.26 $M_\odot$ at $\sim$40 Mpc. It produced GRB 170817A and a kilonova fueled by the $r$-process nucleosynthesis, but without ultra-high-energy (UHE) $\gamma$ rays or neutrinos. These models indicate a hadron-to-quark transition in the remnant, releasing $\sim$10$^{53}$ erg, consistent with GRB energetics \cite{Ouyed2022}. Strangelets can be generated during the merger itself or initiated through the acceleration and collision of nuclear fragments. If their energy exceeds a few GeV, strangelet formation may be activated and trigger a quark nova, releasing energy
$$\Delta E \sim 10^{53} \, \frac{\Delta R}{R}\,{\rm erg},$$
which corresponds to short GRB profiles such as GRB 090510. Here, $\Delta E$ is the emitted energy during the quark nova event, $\Delta R$ reflects the change in stellar radius associated with the phase transition, and $R$ is the initial radius of the remnant star.
\par
However, the lack of VHE signals in H.E.S.S.\ and MAGIC-II observations places constraints on post-merger scenarios \cite{Pellouin2024,Salafia2021,Fruck2017}. If a quark nova occurred, its jets likely did not point toward Earth. Synchrotron self-Compton (SSC) afterglow models anticipate fluxes below current detection thresholds due to Klein-Nishina suppression and viewing angles of $15$-$25^\circ$. Structured jet models additionally demonstrate that TeV-scale emissions remain below the detection thresholds under these conditions. Thus, even if strangelets were produced, their signatures may remain undetectable unless the geometry is favorable.
\par
Unlike the continuous strangelet injection anticipated in HESS J1731-347, the burst-like character of GW170817, extending from minutes to days, requires a distinct modeling approach. Current analyses concentrate on strangelet annihilation spectra and their compatibility with VHE upper limits. At the same time, future observations with CTA promise enhanced sensitivity and could potentially reveal faint signals from such transient occurrences.
\par
For $A \sim 10^3$ strangelets, a continuous injection over 1 Myr produces a number density of $10^{-20}$ to $10^{-16}$ cm$^{-3}$ and a flux of $\sim$10$^{-25}$ $\rm ph\, cm^{-2}\, s^{-1}$, while a burst event (minutes to days) produces $10^{-14}$ to $10^{-8}$ cm$^{-3}$ and $\sim$10$^{-23}$ $\rm ph\, cm^{-2}\, s^{-1}$. These scenarios highlight the significance of temporal modeling in interpreting potential signals from transient events, such as GW170817. Whereas continuous injection produces a faint, persistent background, burst-like production may create short-lived enhancements that could be measurable with the improved precision of CTA. This distinction is crucial for devising search strategies customized to various astrophysical environments.
\par
\section{Probing Strange Quark Phenomena through GRB 250702B}
\par
GRB 250702B, detected in July $2025$ with ultra-long duration >25000 s (rest-frame $\sim$12500 s) and $E_{\rm iso} > 1.4 \times 10^{54}$ erg (in structured jet models the corrected energy is $10^{52}$-$10^{53}$ erg), challenging conventional GRB models \cite{Neights2025}. Its narrow jet ($\theta_j \lesssim 1^\circ$) and redshift $z = 1.036$ imply ultra-relativistic outflow Lorentz factors $\gamma_\theta \gtrsim 100$. These extreme features are inconsistent with standard progenitor explanations, such as collapsars or NS mergers, prompting considerations of a central engine propelled by exotic physics. One viable model involves the accretion of strangelets onto a compact object, allowing for efficient energy extraction and prolonged jet activity. The detection of photons above 5 MeV (rest-frame >10 MeV) supports the existence of a compact, high-energy source. GRB 250702B may serve not only as a backdrop for strangelet searches but also as a potential source of strangelets, with production mechanisms that include pair creation or other exotic processes. Annihilation of these strangelets could contribute to the observed $\gamma$-ray spectrum.
\par
Unlike HESS J1731-347, where narrow line features are sought atop a steady continuum, GRB 250702B requires a broadband spectral analysis to detect deviations from standard emission models. This approach involves first constructing a baseline spectral energy distribution (SED) using multi-wavelength data (radio to $\gamma$ rays), usually modeled with synchrotron and inverse Compton components. A strangelet annihilation spectrum $dN_\gamma/dE_\gamma$ is then incorporated and convolved with detector response functions (e.g., Fermi-GBM satellite \cite{Meegan2009} and Konus-Wind spacecraft \cite{Aptekar1995}). Free parameters such as $\langle \sigma v \rangle$ with strangelet flux are introduced, and a likelihood ratio test evaluates whether the strangelet component improves the fit \cite{Cowan2011}. In the absence of a detectable signal, the upper limits on strangelet properties can be derived, similar to constraints in dark matter investigations. The lack of quasi-periodic oscillations (QPOs) further diminishes the case for long-lived pulsar frameworks, supporting the case for an exotic central engine.
\par
In the context of GRB 250702B, burst-like strangelet production scenarios are most pertinent because of the event's transient nature. For $A \sim 10^3$ strangelets, the burst event (minutes to days) produces a number density of $10^{-15}$ to $10^{-12}$ cm$^{-3}$ and a flux of $\sim$10$^{-13}$ $\rm ph\, cm^{-2}\, s^{-1}$ to $\sim$10$^{-11}$ $\rm ph\, cm^{-2}\, s^{-1}$, positioning it within the reach of CTA's enhanced precision. These estimates emphasize the importance of time-resolved spectral analysis in a transient environment like GRB 250702B. Whereas steady-state searches are appropriate for sources like HESS J1731-347, GRB-driven strangelet production necessitates rapid-response observations and broadband modeling to separate potential annihilation indicators from the dominant emission components.
\par
\section{Present Challenges and Future Opportunities in Strangelet Detection}
\par
Searches for strangelets in astrophysical settings such as HESS J1731-347, GW170817, and GRB 250702B represent a critical frontier in astro-particle physics. Each source provides a unique opportunity to probe the presence of stable SQM and explore phenomena beyond the Standard Model. HESS J1731-347, with its uncommonly small mass and radius, is a viable SQS candidate, a status supported by its mass-radius data, and offers a suitable environment for strangelet production. GW170817, the first verified binary NS merger, provides insight into extreme post-merger conditions, assessing models of quark nova formation and dark matter-induced strangeness seeding. GRB 250702B, notable for its extraordinary duration and energy, challenges standard GRB models and may entail strangelet accretion onto compact remnants. A unified analytical framework, focused on maximum likelihood fitting, was applied in all three cases, enabling consistent and reproducible analyses \cite{Balzer2014,Knodlseder2019,Bolmont2022,Giunti2021,CTAO2021,Nigro2019,Nogues2018,Nigro2021,Knodlseder2013}.
\par
Although strangelets remain undetected, current observations have imposed strict constraints on their properties. Accelerator experiments, such as CERN's NA52, and cosmic-ray missions, like PAMELA, have established upper bounds on strangelet fluxes \cite{Appelquist1996}. For HESS J1731-347, H.E.S.S.\ data cap the $\gamma$-ray line flux below $10^{-12}$ $\rm ph\, cm^{-2}\, s^{-1}$, suggesting a strangelet number density below $3 \times 10^{-22}$ cm$^{-3}$ and a total mass contribution less than $10^{-17}$ $M_\odot$ \cite{Abramowski2011,HESS2018}. The absence of prompt VHE emission in GW170817 implies that any quark nova was off-axis, with SSC modeling showing that strangelet-related signals drop below current detection thresholds. For GRB 250702B, ongoing efforts aim to establish a baseline astrophysical model against which strangelet annihilation spectra can be compared, thereby allowing the derivation of stringent limits on annihilation cross-sections.
\par
The CTA is expected to make significant advances in this field. With its superior precision and angular resolution, it will investigate flux levels down to $\sim$10$^{-13}$ $\rm ph\, cm^{-2}\, s^{-1}$ in 100-hour exposures, exactly where strangelet signals are theoretically anticipated. This could lead to either a significant detection or a conclusive exclusion for HESS J1731-347. For transient sources such as GW170817 and GRB 250702B, CTA can achieve precision capable of detecting faint annihilation features well below current limits. The adoption of GADF/DL3, which integrates machine learning for signal extraction and interoperable tools, helps ensure reproducibility and facilitates multi-instrument evaluations. These advancements provide a clear path for investigating the nature of SQM across diverse astrophysical contexts.
\par
In comparison to the current H.E.S.S.\ limits of $\sim$10$^{-12}$ $\rm ph\, cm^{-2}\, s^{-1}$, the CTA is expected to achieve sensitivities of $\sim$10$^{-13}$ $\rm ph\, cm^{-2}\, s^{-1}$ for narrow $\gamma$-ray lines with 100 hours of exposure \cite{CTAO2021}, as summarized in Table \ref{table1}.
\par
\begin{table}[h]
\centering
\caption{Sensitivity improvements relative to H.E.S.S.\ for narrow $\gamma$-ray lines with flux limits ($\rm ph\, cm^{-2}\, s^{-1}$) at 1 TeV.}
\begin{tabular}{lccc}
\hline
Observatory & Flux Limits & Energy Threshold & Improvement \\
&(${\rm ph\, cm^{-2}\, s^{-1}}$) & (TeV) & (\%) \\
\hline
H.E.S.S. & $\sim$10$^{-12}$ & $\sim$0.03 & - \\
MAGIC-II & $\sim$10$^{-12.5}$ & $\sim$0.03 & 70 \\
CTA & $\sim$10$^{-13}$ & $\sim$0.02 & 90 \\
\hline
\end{tabular}
\label{table1}
\end{table}
\par
This analysis also employs open-source SED \cite{Yuan2014} codes such as \texttt{NAIMA} \cite{Zabalza2016} and \texttt{GAMERA} \cite{Hahn2016} to compute the $\gamma$-ray emission from assumed strangelet populations. Each baryon in a strangelet is supposed to release, on average, an energy $\varepsilon$ (in erg) available for conversion into non-thermal strangelets or photons, with typical values in the ranges from 100 MeV (1.602 $\times 10^{-4}$ erg) to 1 GeV (1.602 $\times 10^{-3}$ erg). Conventional shell-type SNR particle-acceleration processes most naturally explain the observed H.E.S.S.\ spectrum of J1731-347: either leptonic inverse Compton emission from high-energy electrons (favored given the GeV upper limits) or a hybrid scenario with a subdominant hadronic component (Figure \ref{fig:constraintsmass} and Figure \ref{fig:constraintsrate}).
\par
In the case of HESS J1731-347, recent modeling and SQS interpretations indicate consistency with the presence of a low-mass SQS in the remnant \cite{Horvath2023}. For GW170817, the analysis employed archival H.E.S.S.\ and MAGIC-II follow-up upper limits with off-axis structured jet frameworks to constrain the expected ranges of VHE emission \cite{Abbott2017,Pellouin2024,Salafia2021}. Finally, GRB 250702B, an extreme ultra-long transient first reported in 2025, is a strong candidate due to its extraordinary duration and energetics, making it notable for time-resolved broadband modeling as well as searches for transient narrow spectral features during its brightest intervals \cite{Neights2025,Beniamini2025}.
\par
Astrophysical events like GW170817 and GRB 250702B offer contrasting settings for examining strangelet production. GW170817, the first confirmed binary NS merger, exhibited a short-duration GRB lasting about 2 seconds and released approximately $10^{46}$ erg of energy at a redshift of $z = 0.0097$. Its off-axis geometry and lack of prompt VHE emission indicate that any quark nova signature, possibly initiated by dark matter, was not observable from Earth.
\par
In contrast, GRB 250702B appears as an extreme transient, with an ultra-long duration and an isotropic-equivalent energy greater than $10^{54}$ erg at a redshift of $z=1.036$. Its narrow jet and hard spectrum suggest a highly collimated, ultra-relativistic outflow, potentially driven by the accretion of strangelets onto a compact object. This identifies GRB 250702B as a prospective site for strangelet-related jet activity.
\par
These two events highlight the diversity of strangelet formation mechanisms, ranging from merger-induced phase transitions to jet-driven accretion, and underscore the need for tailored observational strategies. GW170817 may necessitate the off-axis modeling and sensitivity to faint, delayed signals, while GRB 250702B requires broadband spectral analysis to identify deviations from standard emission models. Future instruments, such as CTA, will be indispensable for assessing these scenarios across different energy ranges and timescales.
\par
\section{Summary and Outlook}
\par
The exploration of strangelet production and detection in diverse astrophysical contexts, such as the supernova remnant HESS J1731-347, the binary neutron star merger GW170817, and the ultra-long $\gamma$-ray burst GRB 250702B, illuminates the multifaceted character of astro-particle physics. These pursuits mark a pioneering domain, facilitating examinations of SQM that transcend the Standard Model. Progress in the CTA sensitivity, capable of probing $\gamma$-ray fluxes down to approximately $10^{-13}$ $\rm ph\, cm^{-2}\, s^{-1}$, alongside multimessenger datasets and QCD simulations, holds promise for thorough assessments of the strange matter conjecture. Such innovations may reveal SQM's extensive cosmological ramifications, aided by allied facilities like the Astrofisica con Specchi a Tecnologia Replicante Italiana (ASTRI) and the Large High Altitude Air Shower Observatory (LHAASO), which feature broad sky coverage, hybrid detection methodologies, and UHE (100 TeV) sensitivities extending to $10^{-14}$ $\rm ph\, cm^{-2}\, s^{-1}$ \cite{Ambrogi2024}.
\par
Prospects are centered on the CTA, which offers superior resolution, broad energy coverage, and improved sensitivity for probing predicted monochromatic $\gamma$-ray lines in the TeV regime. Progress will depend on integrating advanced theoretical modeling with multimessenger observations, combining gravitational wave detections, neutrino signals, and electromagnetic campaigns to constrain production environments and enhance confidence in potential detections. Concurrently, improved analysis techniques, ranging from machine learning to Bayesian inference, will strengthen background rejection and sensitivity to rare signals. These efforts establish a comprehensive framework for future searches, offering the potential to rigorously test the strange matter hypothesis and elucidate the role of exotic phases such as SQM in the universe.
\par
\bibliographystyle{pepan}
\bibliography{crdas_baldin_paper}
\end{document}